\documentclass[sigconf,balance=true]{acmart}

\AtBeginDocument{%
  }

\usepackage{algorithm}
\usepackage{algorithmic}
\usepackage{subcaption}
\usepackage{multirow}
\usepackage{colortbl}
\usepackage{xspace}
\usepackage{pifont}
\usepackage{enumitem}

\makeatletter
\pretocmd{\@mktitle}{%
  \let\@plaintitle\@title
  \def\@title{%
    \raisebox{-0.3\height}{\includegraphics[height=2.3em]{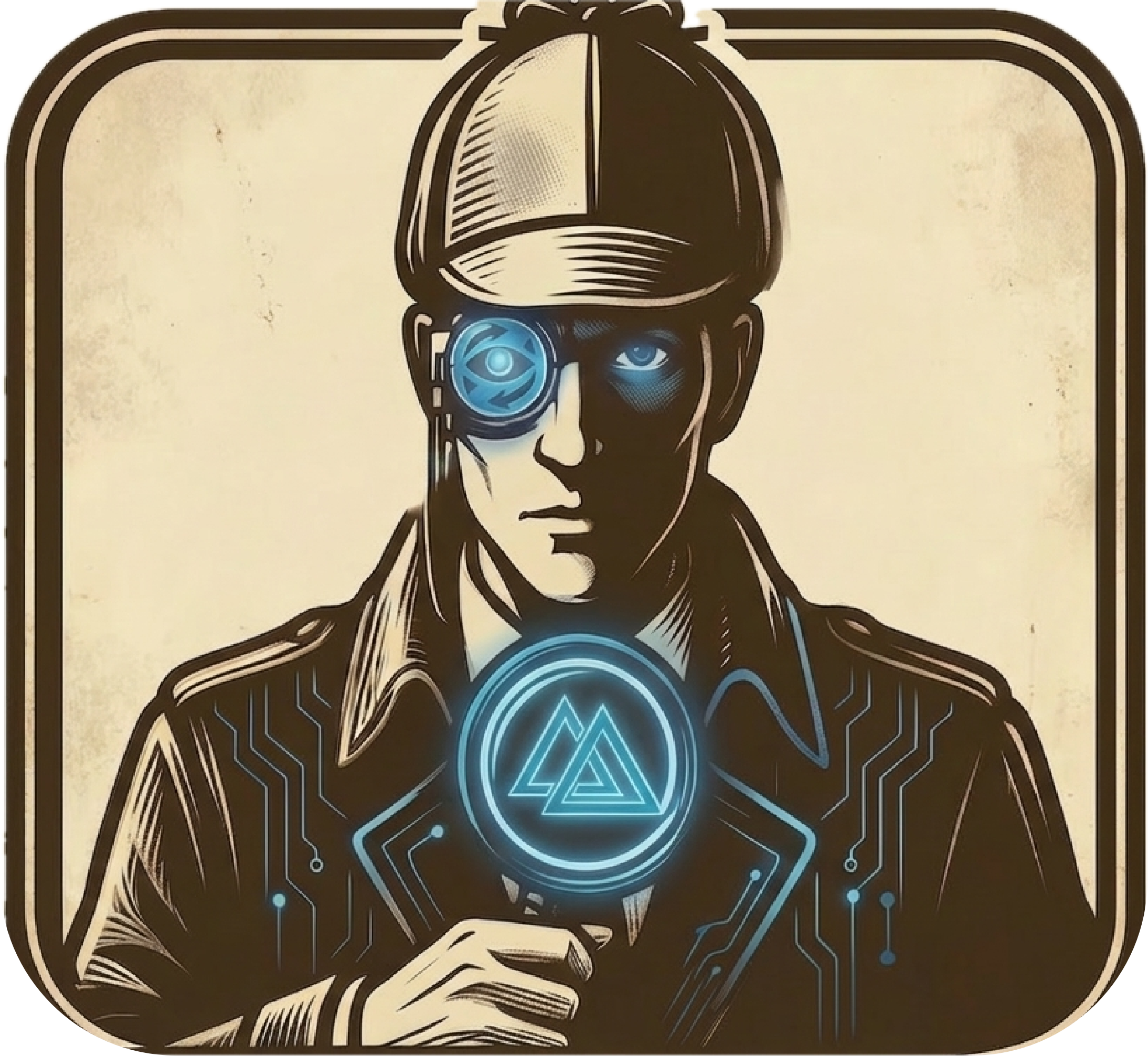}}%
    \hspace{0.5em}%
    \parbox[c]{0.85\textwidth}{\@plaintitle}%
  }%
}{}{}
\apptocmd{\@mktitle}{\let\@title\@plaintitle}{}{}
\makeatother

\definecolor{bestbg}{HTML}{E8F5E9}   
\definecolor{ablationbg}{HTML}{F3E5F5} 

\newcommand{\method}{EvoSherlock\xspace}
\newcommand{\task}{$L^2$-SCE\xspace}
\newcommand{\doE}{$do(E)$\xspace}
\newcommand{\doM}{$do(M)$\xspace}
\newcommand{\doT}{$do(T)$\xspace}
\newcommand{\ie}{\textit{i.e.},\xspace}
\newcommand{\eg}{\textit{e.g.},\xspace}
\newcommand{\etal}{\textit{et al.}\xspace}
\newcommand{\vs}{\textit{vs.}\xspace}

\copyrightyear{2026}
\acmYear{2026}
\setcopyright{cc}
\setcctype{by}
\acmConference[MM '26]{Proceedings of the 34th ACM International Conference on Multimedia}{November 10--14, 2026}{Rio de Janeiro, Brazil}
\acmBooktitle{Proceedings of the 34th ACM International Conference on Multimedia (MM '26), November 10--14, 2026, Rio de Janeiro, Brazil}
\acmDOI{10.1145/3767308.3836159}
\acmISBN{979-8-4007-2213-4/2026/11}

\begin{document}

\title[EvoSherlock: Towards Agentic Lifelong Evolution for Unseen Long-Tailed Security-Critical Events in Videos]{EvoSherlock: Towards Agentic Lifelong Evolution for Unseen Long-Tailed Security-Critical Events in Videos}


\author{Zixin Fan}
\orcid{0009-0003-5316-0630}
\affiliation[obeypunctuation=true]{%
  \institution{School of Computer Science and Technology, Soochow University},
  \city{Suzhou},
  \country{China}
}
\email{20255227111@stu.suda.edu.cn}

\author{Jiahong Lu}
\orcid{0009-0009-2409-5511}
\affiliation[obeypunctuation=true]{%
  \institution{School of Computer Science and Technology, Soochow University},
  \city{Suzhou},
  \country{China}
}
\email{20255227109@stu.suda.edu.cn}

\author{Changsheng Zheng}
\orcid{0009-0009-7182-5759}
\affiliation[obeypunctuation=true]{%
  \institution{School of Cyber Science and Engineering, Qufu Normal University},
  \city{Qufu},
  \country{China}
}
\email{changshengzheng@qfnu.edu.cn}

\author{Yu Hong}
\orcid{0000-0003-0606-3718}
\affiliation[obeypunctuation=true]{%
  \institution{School of Computer Science and Technology, Soochow University},
  \city{Suzhou},
  \country{China}
}
\email{tianxianer@gmail.com}

\author{Jingjing Wang}
\authornote{Corresponding Author: Jingjing Wang.}
\orcid{0009-0006-3619-1525}
\affiliation[obeypunctuation=true]{%
  \institution{School of Computer Science and Technology, Soochow University},
  \city{Suzhou},
  \country{China}
}
\email{djingwang@suda.edu.cn}

\begin{abstract}
Existing Security-oriented Video Understanding (SVU) systems assume a \emph{closed world}, \ie static category sets, abundant labels, and the premise that all event types are known upfront. Real-world security-critical events break these assumptions: they follow long-tailed distributions, new types emerge continuously, and critical security events may offer only a few samples. We formalize this gap as \textbf{Lifelong Evolving Task for Long-Tailed Security-Critical Events in Videos ({\boldmath$L^2$}-SCE)}, a new task that requires VLMs to continually classify and temporally localize newly emerging security-critical events from scarce samples without forgetting previously learned events. Furthermore, \task reveals two critical challenges: (1)~\textbf{Intra-Event Scarcity}, where extreme data scarcity may weaken both classification and temporal localization for new events, and (2)~\textbf{Inter-Event Interference}, where cross-event feature entanglement and representation drift may strengthen catastrophic forgetting. On this basis, we propose \textbf{\method}, a causal-enhanced approach orchestrated end-to-end by an \textbf{Agentic Controller} with self-reflective closed-loop control, which includes two core modules: the Intra-Event \textbf{C}ausal \textbf{V}ideo \textbf{G}eneration module (\textbf{CVG}) and the Inter-Event \textbf{C}ausal \textbf{D}ecoupling and \textbf{A}lignment module (\textbf{CDA}), to address the above two challenges, respectively. Especially, this paper constructs a \task dataset to simulate real-world incremental conditions. Extensive experiments on our benchmark demonstrate the advantages of \method over several advanced baselines. These justify the importance of the proposed \task and the effectiveness of \method in classifying and temporally localizing emerging security-critical events from scarce samples.
\end{abstract}



\begin{teaserfigure}
  \centering
  \includegraphics[width=\textwidth]{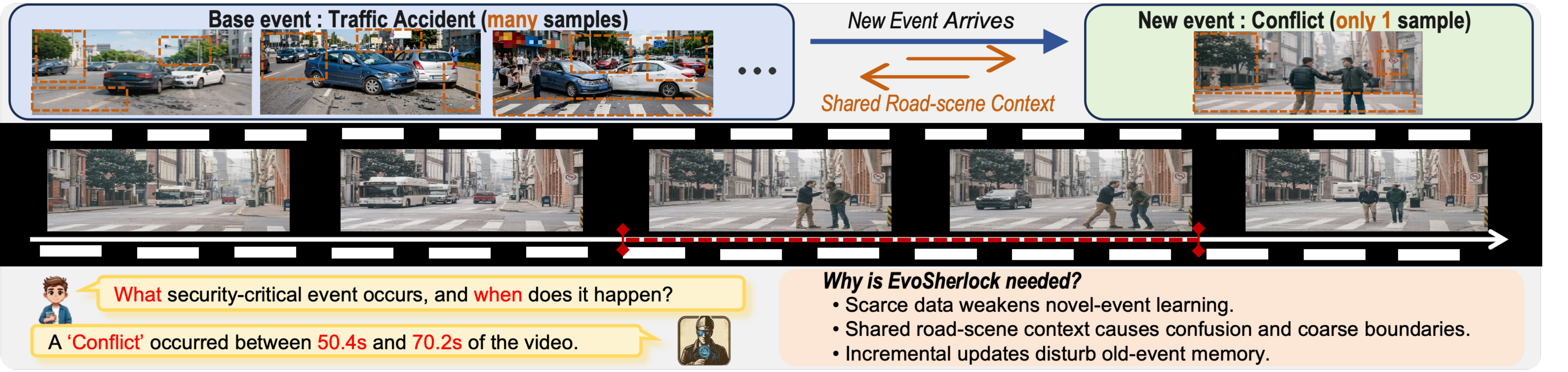}
  \caption{Overview of \textbf{{\boldmath$L^2$}-SCE} and its challenges addressed by \method. ``Conflict,'' with one training sample, shares context with the base event ``Traffic Accident,'' requiring classification and temporal localization without forgetting prior events.}
  \Description{An example figure for Lifelong Evolving Task for Long-Tailed Security-Critical Events in Videos. A new class ``Conflict'' arrives with only one training sample and shares road-scene context with the base event (i.e., previously learned event ``Traffic Accident''), requiring the model to classify the event and localize its temporal extent without forgetting previously learned events.}
  \label{fig:teaser}
\end{teaserfigure}

\maketitle

\section{Introduction}
\label{sec:intro}

Real-world Security-oriented Video Understanding (SVU)~\cite{jin2026deepsvu,liu2024generalized,wu2024deep,yuan2024uca} is an open-ended problem, in stark contrast to the \emph{closed-world} assumption of static event sets with abundant supervision. Despite rapid progress~\cite{pang2021deep,sultani2018real,radford2021learning,wu2024vadclip,joo2023cliptsa,chen2024internvl,zanella2024harnessing}, current methods assume fixed event sets and sufficient data, whereas real-world security-critical events are long-tailed and continually emerging: common incidents (\eg traffic accidents) offer abundant samples, while rare yet critical events (\eg drowning) may provide only a few~\cite{wang2024comprehensive}. Data scarcity, quality-controlled synthesis, and catastrophic forgetting compound across events, making them intractable for any single static model, calling for an agentic paradigm~\cite{zheng2025lifelong,wang2024survey_agent} that orchestrates perception, generation, and self-correction across events.

To formalize this gap, we propose \textbf{Lifelong Evolving Task for Long-Tailed Security-Critical Events in Videos ({\boldmath$L^2$}-SCE)}, a task addressing long-tailed security-critical event classification and temporal localization under an event-incremental lifelong learning setting. Security-critical events arrive sequentially with highly skewed sample counts; for each new event, the model must learn its category and temporal boundaries $[\hat{t}_s,\hat{t}_e]$ from scarce samples while retaining all previously observed events, ideally with non-negative backward transfer. Figure~\ref{fig:teaser} illustrates this requirement when the one-sample ``Conflict'' event visually overlaps with the previously learned ``Traffic Accident.'' \task therefore faces two major challenges, which are detailed as follows.

On the one hand, learning new security-critical events from extremely scarce samples while preserving precise temporal localization is a core challenge we call \textbf{Intra-Event Scarcity}. With as little as one example, shared environmental cues can cause event confusion and imprecise temporal boundaries~\cite{zhang2022actionformer,shi2023tridet,wang2024temporal}. Conventional augmentation changes only low-level appearance, while unconstrained video generators~\cite{wan2025wan,brooks2024video,blattmann2023svd} may introduce semantic hallucinations and temporal misalignment~\cite{zanella2024harnessing}. Such failures can alter action semantics or event timing, injecting label noise into classification and temporal boundary supervision. Therefore, a well-designed approach to \task should be able to synthesize causally faithful training samples that respect the underlying causal mechanisms of events, preserving both event semantics and precise temporal boundaries, thereby effectively addressing the Intra-Event Scarcity challenge caused by extreme data scarcity.

On the other hand, catastrophic forgetting from cross-event feature entanglement and representation drift is an equally serious problem we call \textbf{Inter-Event Interference}. In \task, a shared LoRA adapter~\cite{hu2022lora,han2024parameter} allows new-event updates to overwrite features important to old events~\cite{kirkpatrick2017overcoming}. Shared contexts further exacerbate event confusion, while the evolving feature extractor makes stored prototypes increasingly stale~\cite{de2021continual}. This mismatch compounds retrieval errors because stored prototypes no longer align with the current embedding space. Existing strategies based on replay, regularization, or architectural adaptation~\cite{rebuffi2017icarl,lopez2017gradient,kirkpatrick2017overcoming,li2018learning,wang2024comprehensive,zhou2024class} mitigate forgetting but do not separate causal mechanisms from environmental confounders. Therefore, a well-designed approach to \task should disentangle invariant causal mechanism features from contextual confounders and maintain prototype consistency as the embedding space evolves, thereby effectively mitigating the Inter-Event Interference problem.

To tackle these challenges, we propose \textbf{\method}, a causal-enhanced agentic approach for \task. To address Intra-Event Scarcity, its \textbf{Intra-Event Causal Video Generation (CVG)} module generates causally grounded counterfactual samples with group-level quality gating. For Inter-Event Interference, its \textbf{Inter-Event Causal Decoupling and Alignment (CDA)} module separates contextual confounders from causal mechanisms and maintains prototype consistency. An \textbf{Agentic Controller} orchestrates the pipeline through self-reflection and quality-gated updates, and we construct a \task dataset for evaluation. Experiments show that \method outperforms all baselines on classification, temporal localization, and lifelong stability, achieving reduced forgetting and positive backward transfer, confirming that both CVG and CDA are necessary for \task.

\section{Related Work}
\label{sec:related}

\subsection{Security-oriented Video Understanding}
\label{sec:rw_svu}

Security-oriented Video Understanding (SVU) detects, localizes, and interprets security risks in surveillance videos~\cite{jin2026deepsvu,liu2024generalized,wu2024deep}. Early methods~\cite{pang2021deep,sultani2018real,tian2021rtfm} are limited to binary event scoring without category semantics. Visual-language alignment via CLIP~\cite{radford2021learning,joo2023cliptsa,wu2024vadclip} and training-free approaches~\cite{zanella2024harnessing,wu2024open} enable semantically diverse detection in a zero-shot setting. Recent methods~\cite{li2025anomize,zhang2024holmesvad,zhao2024hawkeye,huang2025exvad,jin2026deepsvu,ma2025skynet,ma2025sherlock,luo2025omnisila,ouyang2025conan} and multimodal benchmarks~\cite{yuan2024uca,liu2025surveillancevqa} further push SVU toward fine-grained explainability. Nevertheless, all existing paradigms presuppose a fixed category space with abundant per-event samples, without addressing long-tailed distributions or continual category integration. \textbf{To bridge this gap}, we introduce \task, the first long-tailed security-critical event classification and temporal localization task under an event-incremental setting, and propose \method.

\subsection{Agentic Lifelong Learning}
\label{sec:rw_fscil}

Lifelong learning embodies the stability-plasticity dilemma~\cite{wang2024comprehensive,chen2018lifelong}. Classical approaches address catastrophic forgetting through regularization, distillation, replay, and bias correction~\cite{kirkpatrick2017overcoming,li2018learning,rebuffi2017icarl,buzzega2020dark,wu2019large}, but all assume sufficient per-event samples. Few-shot class-incremental methods~\cite{zhou2022forward,peng2022alice,song2023savc,yang2023ncfscil,wang2023teen} reserve representational capacity via geometric constraints or prototype recalibration~\cite{de2021continual}, while generative replay~\cite{shin2017continual,wu2018memory,ostapenko2019learning,gao2023ddgr} synthesizes pseudo-samples but lacks group-level temporal consistency. Lifelong learning video benchmarks~\cite{villa2022vclimb,tang2024vilco} and methods such as Qin \etal~\cite{ccgs2024} target action recognition but omit temporal security event localization, leaving \task unaddressed. Recently, LLM-based agents~\cite{zheng2025lifelong,wang2024survey_agent,wang2023voyager,wang2026embodied} advance lifelong learning toward autonomous perception-memory-action loops, and Yang \etal\cite{yang2025panda} adopt an agentic pipeline for video security-critical event detection but operate in a training-free zero-sample paradigm without incremental updates. \textbf{In contrast}, \method introduces a closed-loop agentic framework unifying causal generation, quality-gated verification, and incremental adaptation, enabling continual assimilation of new event types for \task.

\subsection{Causal Visual Understanding}
\label{sec:rw_causal}

Causal reasoning addresses spurious correlations in visual understanding. Arjovsky \etal\cite{arjovsky2019invariant} and Sch\"olkopf \etal\cite{scholkopf2021toward} formalize environment-invariant feature extraction via structural causal models. In practice, $do$-calculus has been applied to scene graph debiasing, causal attention, and vision-language alignment~\cite{tang2020unbiased,wang2021causal,yang2021catt}, while counterfactual interventions~\cite{yue2021counterfactual,liu2021counterfactual} generate augmented samples respecting causal structure. Diffusion and flow-matching generators~\cite{wan2025wan,brooks2024video,blattmann2023svd} enable video-level counterfactual synthesis, though quality-gated curation outperforms naive scaling~\cite{blattmann2023svd}. Crucially, existing causal visual methods apply corrections at inference time with a fixed feature extractor; none maintain causal consistency across incremental events with evolving representations. \textbf{Unlike these methods}, \method embeds causal intervention into both the forward pass and data synthesis, enabling lifelong causal decoupling.

\section{Method}
\label{sec:method}

\begin{figure*}[t]
  \centering
      \includegraphics[width=\textwidth]{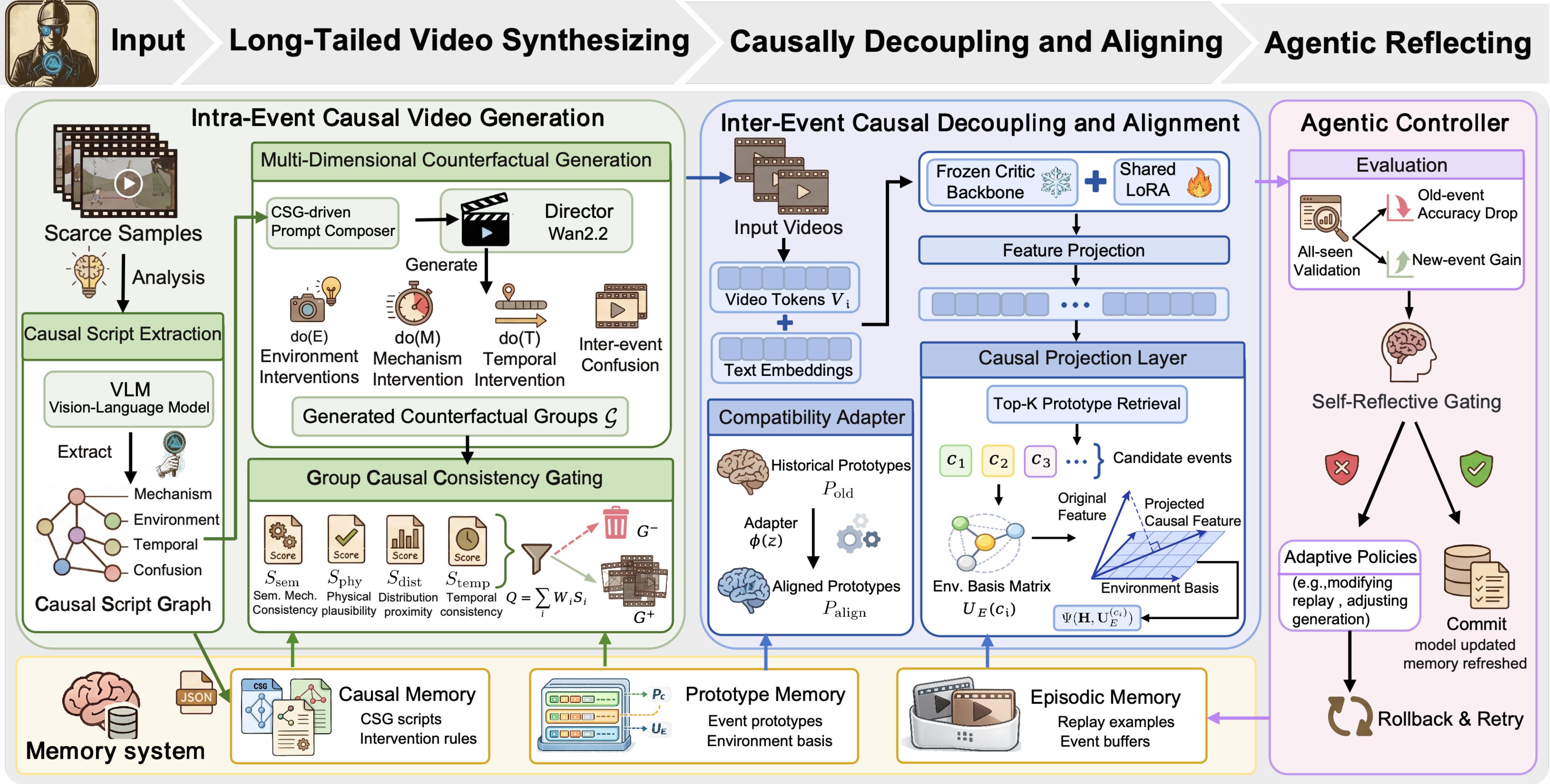}
  \caption{Overview of \method. (a)~Intra-Event CVG: extracts CSGs from scarce samples and synthesizes counterfactual videos along \doE, \doM, \doT, and confusion axes with GCCG quality gating. (b)~Inter-Event CDA: decouples causal features from environmental confounders and realigns drifted prototypes. (c)~Agentic Controller: self-reflective gating with adaptive policies or rollback. (d)~Memory System: stores CSG scripts, event prototypes, and replay exemplars.}
  \Description{Architecture diagram of EvoSherlock showing a four-stage pipeline (Input and Analysis, Generation Curriculum, Feature Processing and Training, Agentic Control) with a shared Memory System comprising causal, prototype, and episodic stores.}
  \label{fig:overview}
\end{figure*}

As discussed in Section~\ref{sec:intro}, \task faces two key challenges: \emph{Intra-Event Scarcity} and \emph{Inter-Event Interference}. \method addresses them with three components (Figure~\ref{fig:overview}): (1)~\textbf{Intra-Event Causal Video Generation (CVG)} (Section~\ref{sec:csg-acc}), (2)~\textbf{Inter-Event Causal Decoupling and Alignment (CDA)} (Section~\ref{sec:cp-cc}), and (3)~an \textbf{Agentic Controller} (Section~\ref{sec:controller}) for closed-loop orchestration. Given scarce samples $D_{\text{new}}^c$, model $f_\theta$, and memory $\mathcal{B}$, \method produces updated $f_{\theta'}$ with predictions $(\hat{y}, \hat{I})$ (Algorithm~\ref{alg:pipeline}).

\subsection{Problem Definition and Setup}
\label{sec:problem}

\subsubsection{Task Formulation}
\label{sec:task_formulation}

\task is defined as follows. Let $\mathcal{C}$ denote the set of security-critical event types and $V$ a video. The model outputs a classification $\hat{y}\!\in\!\mathcal{C}$ and a temporal interval $\hat{I}\!=\![\hat{t}_s, \hat{t}_e]$. The events are divided into base events $\mathcal{C}_0$ and incremental events $\mathcal{C}_1, \dots, \mathcal{C}_T$, whose sample counts follow a heavily skewed long-tailed distribution. In event $t$, $\mathcal{C}_t$ arrives with extremely scarce training samples; we denote the scarce training set for a specific event $c \in \mathcal{C}_t$ as $D_{\text{new}}^c$. The model is updated using $D_{\text{new}}^c$ together with a bounded episodic memory $\mathcal{B}$ (a small set of stored exemplars from previously seen events), and evaluated on $\bigcup_{i=0}^{t}\mathcal{C}_i$, with the aim of continually classifying and temporally localizing security-critical events without forgetting previously learned events.

\subsubsection{Backbone}
\label{sec:backbone}

We employ Qwen3-VL-8B-Instruct~\cite{bai2025qwen3vl} as the backbone, fine-tuned with a single shared LoRA~\cite{hu2022lora} ($r\!=\!16$, $\alpha\!=\!32$) while keeping the main parameters frozen. Two task heads are attached: a classification head that outputs $\hat{y}\!\in\!\mathcal{C}$ and a temporal regression head that predicts $\hat{I}\!=\![\hat{t}_s, \hat{t}_e]$. This shared-adapter design reduces overhead but induces inter-event interference, which CDA addresses (Section~\ref{sec:cp-cc}).

\subsection{CVG: Intra-Event Causal Video Generation}
\label{sec:csg-acc}

To address the Intra-Event Scarcity challenge, we design the CVG module with three blocks: Causal Script Graph~(CSG) Extraction, Multi-Dimensional Counterfactual Generation, and Group Causal Consistency Gating~(GCCG).

\subsubsection{Causal Script Graph (CSG) Extraction}
\label{sec:csg}

To ensure generated samples are causally faithful, the generator must first formalize what constitutes each event. Given scarce training samples $D_{\text{new}}^c$ for a new event $c$, a VLM automatically extracts a structured \emph{Causal Script Graph}
{$\text{CSG}^{(c)}\!=\!(\mathcal{M},\!\mathcal{E},\!\mathcal{T},\!\mathcal{C}_{\text{conf}},\!\mathcal{I})$}
with five fields:

\noindent\textbf{(1)~Mechanism} $\mathcal{M}$: core causal event, ordered sub-events, necessary conditions (whose absence defines near-miss negatives), duration range $[\tau_{\min}, \tau_{\max}]$, and intensity factors.

\noindent\textbf{(2)~Environment} $\mathcal{E}$: scene, camera, and nuisance variables (\eg lighting, crowd density) that change appearance but do not classify events.

\noindent\textbf{(3)~Temporal Anchors} $\mathcal{T}$: evidence timestamps and observable start/end cues.

\noindent\textbf{(4)~Confusable Events} $\mathcal{C}_{\text{conf}}$: top-$K_c$ ($K_c\!=\!3$) visually or semantically similar existing events retrieved from prototype memory.

\noindent\textbf{(5)~ Intervention Set} $\mathcal{I}$: concrete operations \doE, \doM, and \doT applicable to this event, derived from $\mathcal{M}$, $\mathcal{E}$, and $\mathcal{T}$.

\noindent The CSG guides generation and gating, and is stored in causal memory for later refinement.

\subsubsection{Multi-Dimensional Counterfactual Generation}
\label{sec:generation}

Given the formalized causal structure, we generate along four complementary axes to maximize diversity while preserving discriminability. Taking the CSG and the original samples $D_{\text{new}}^c$ as input, a Director (Wan2.2~\cite{wan2025wan} video generator guided by CSG prompts) produces a raw synthetic set $G_{\text{raw}}$ as follows:

\noindent\textbf{(1)~}{\boldmath$do(E)$} randomizes scene, lighting, and camera while preserving the mechanism and temporal boundaries; the resulting feature residuals are also used to estimate the environment basis $\mathbf{U}_E^{(c)}$ (Section~\ref{sec:projection}).

\noindent\textbf{(2)~}{\boldmath$do(M)$} alters key mechanism elements (\eg action speed, removing a necessary condition from $\mathcal{M}$) to produce near-miss negatives with similar temporal profiles but without the causal mechanism.

\noindent\textbf{(3)~}{\boldmath$do(T)$} shifts event onset or duration by a random temporal offset $\Delta$ (sampled uniformly from the valid range), providing equivariance supervision for the temporal head.

\noindent\textbf{(4)~Confusion} samples resemble confusable old events while preserving the new-event mechanism, sharpening event boundaries.

\subsubsection{Group Causal Consistency Gating (GCCG)}
\label{sec:gccg}

GCCG scores each synthetic sample within its \emph{group} of causally related variants using four measures:

\noindent\textbf{Semantic mechanism consistency} $S_{\text{sem}}\!\in\![0,1]$: the VLM checks whether the video matches the CSG's core event and necessary conditions.

\noindent\textbf{Physical plausibility} $S_{\text{phy}}\!\in\![0,1]$: rule-based checks on temporal ordering, duration validity, and inter-frame consistency.

\noindent\textbf{Distribution proximity} $S_{\text{dist}}\!\in\![0,1]$: we isolate the mechanism signal by subtracting the context embedding from the event embedding. Given $\mathbf{z}_{\text{evt}} = f_\theta(V_{[t_s, t_e]})$ and
$\mathbf{z}_{\text{ctx}} = f_\theta(V_{[t_s-\delta_{\text{ctx}},\, t_s) \cup (t_e,\, t_e+\delta_{\text{ctx}}]})$ where $\delta_{\text{ctx}}$ is the context window duration, the mechanism residual is:
\begin{equation}
\mathbf{z}_{\text{mech}} = \frac{\mathbf{z}_{\text{evt}} - \mathbf{z}_{\text{ctx}}}{\|\mathbf{z}_{\text{evt}} - \mathbf{z}_{\text{ctx}}\|_2}
\label{eq:zmech}
\end{equation}
$S_{\text{dist}}$ is the cosine similarity between $\mathbf{z}_{\text{mech}}$ and the event prototype $\mathbf{p}_{\text{mech}}$: $S_{\text{dist}} = \tfrac{\mathbf{z}_{\text{mech}}^\top \mathbf{p}_{\text{mech}}}{\|\mathbf{z}_{\text{mech}}\| \|\mathbf{p}_{\text{mech}}\|}$.

\noindent\textbf{Temporal consistency} $S_{\text{temp}}\!\in\![0,1]$: we build $K_v\!=\!5$ views by varying sampling offset, frame rate, and segment length, then measure boundary agreement:
\begin{equation}
S_{\text{temp}} = \overline{\text{tIoU}} - \gamma(\sigma_s + \sigma_e)
\label{eq:stemp}
\end{equation}
where $\overline{\text{tIoU}}$ is the mean pairwise temporal IoU across $K_v$ views, $\sigma_s, \sigma_e$ are the standard deviations of predicted start and end boundaries, and $\gamma$ is a penalty coefficient controlling the variance penalty.

The four measures yield a composite score with weights $w_1$--$w_4$ (see Section~\ref{sec:impl}):
\begin{equation}
Q = w_1 S_{\text{sem}} + w_2 S_{\text{phy}} + w_3 S_{\text{dist}} + w_4 S_{\text{temp}}
\label{eq:quality}
\end{equation}

\textbf{Group-level consistency} requires \doE groups to maintain stable labels, \doM groups to show directional changes, and near-miss pairs to exhibit high temporal overlap with flipped labels.

Samples that pass the quality threshold $\tau_Q$ enter the training pool $G^+$; those failing only $S_{\text{temp}}$ are \emph{demoted} to classification-only supervision (temporal regression gating in Section~\ref{sec:training}) to avoid corrupting the localization head with noisy temporal labels. The filtered set $G^+$ is merged with original samples $D_{\text{new}}^c$ and episodic memory $\mathcal{B}$ to form the training pool for the subsequent adaptation stage.

\subsection{CDA: Inter-Event Causal Decoupling and Alignment}
\label{sec:cp-cc}

While CVG addresses data scarcity via augmentation, Inter-Event Interference requires decoupling shared representations. CDA achieves this with two blocks: an operator-level Causal Projection Layer and a Compatibility Adapter.

\subsubsection{Operator-Level Causal Projection Layer}
\label{sec:projection}

If environmental features shared across events can be separated from causal mechanisms before the task heads, event-specific updates will not corrupt unrelated representations. We therefore embed a differentiable causal projection operator $\Psi$ in the visual branch. Given the visual feature sequence $\mathbf{H}\!\in\!\mathbb{R}^{T\times D}$:
\begin{equation}
\tilde{\mathbf{H}}_{\text{causal}} = \mathbf{H} - g(\mathbf{H}) \cdot \bigl(\mathbf{H}\, \mathbf{U}_E^{(c)}\bigr) {\mathbf{U}_E^{(c)}}^\top
\label{eq:projection}
\end{equation}
where $\mathbf{U}_E^{(c)}\!\in\!\mathbb{R}^{D\times k}$ is the per-event environment basis for event $c$ (top-$k$ eigenvectors of the \doE feature-residual covariance), and $g(\mathbf{H})\!\in\![0,1]$ is a sigmoid-gated scalar computed by a single linear layer. The resulting operator $\Psi = \mathrm{Id} - g \cdot \mathbf{U}_E^{(c)} {\mathbf{U}_E^{(c)}}^\top$ projects onto $\mathcal{S}_E^\perp$, removing environmental variation; $\tilde{\mathbf{H}}_{\text{causal}}$ is then fed to both task heads.

\textbf{Basis estimation.} $\mathbf{U}_E^{(c)}$ is obtained via SVD on \doE feature residuals $\{\mathbf{H}_{do(E)} - \mathbf{H}_{\text{base}}\}$.

\textbf{Retrieve-then-Project inference.} At test time, we retrieve top-$K_r$ ($K_r\!=\!5$) events from prototype memory, apply $\Psi(\mathbf{H}, \mathbf{U}_E^{(c_i)})$, and select the highest-confidence event.

\subsubsection{Compatibility Adapter}
\label{sec:adapter}

Even with deconfounding, \emph{prototype drift} persists: LoRA updates shift the embedding space, making stored prototypes stale. We design a corrective adapter:
\begin{equation}
\phi(\mathbf{z}) = \operatorname{LayerNorm}(\mathbf{W}\mathbf{z} + \mathbf{b}), \quad \mathbf{W}\!\in\!\mathbb{R}^{D\times D},\; \mathbf{b}\!\in\!\mathbb{R}^D
\label{eq:adapter}
\end{equation}
that maps old-space features into the new causal subspace. Training pairs $(\mathbf{z}_{\text{old}}, \mathbf{z}_{\text{new}})$ come from episodic memory encoded by both the previous and current checkpoints, with $\mathbf{z}_{\text{new}} = \Psi(f_{\theta^*}(x))$. The loss combines MSE and cosine alignment:
\begin{equation}
\mathcal{L}_{\text{comp}} = \|\phi(\mathbf{z}_{\text{old}}) - \mathbf{z}_{\text{new}}\|_2^2 + \lambda_{\text{cos}}\!\left(1 - \frac{\phi(\mathbf{z}_{\text{old}})^\top \mathbf{z}_{\text{new}}}{\|\phi(\mathbf{z}_{\text{old}})\|_2 \|\mathbf{z}_{\text{new}}\|_2}\right)
\label{eq:comp_loss}
\end{equation}
Training occurs after $\mathbf{U}_E^{(c)}$ update, so targets lie in the causal subspace.

\textbf{Causal Invariance Regularization (CIR).}
\label{sec:cir}
While the projection layer deconfounds at inference, the model must also learn during training to be invariant to environmental changes and equivariant to temporal shifts. CIR leverages counterfactual pairs from CVG to impose these constraints:

\noindent\textbf{\doE Invariance.} For environment-intervention pairs $(x, x')$ with the same mechanism:
\begin{equation}
\mathcal{L}_{\text{inv}} = D_{\mathrm{KL}}\!\left(p_\theta(\cdot|x) \| p_\theta(\cdot|x')\right) + \lambda_t \cdot \ell_{\mathrm{SL1}}(\hat{I}, \hat{I}')
\label{eq:inv}
\end{equation}
where $p_\theta(\cdot|x)$ denotes the classification probability distribution over $\mathcal{C}$, $\ell_{\mathrm{SL1}}$ is the Smooth L1 loss, $\hat{I}\!=\![\hat{t}_s, \hat{t}_e]$ the predicted temporal interval, and $\lambda_t$ controls the temporal penalty weight. This enforces invariant classification and temporal predictions under environment changes.

\noindent\textbf{\doT Equivariance.} For temporal-shift pairs $(x, x_{do(T)})$ where $x_{do(T)}$ is obtained by shifting the event onset by $\Delta$:
\begin{equation}
\mathcal{L}_{\text{equiv}} = \ell_{\mathrm{SL1}}(\hat{t}^{do(T)}_s, \hat{t}_s + \Delta) + \ell_{\mathrm{SL1}}(\hat{t}^{do(T)}_e, \hat{t}_e + \Delta)
\label{eq:equiv}
\end{equation}

\noindent\textbf{Orthogonal Protection.} To prevent the projection from removing mechanism features:
\begin{equation}
\mathcal{L}_{\text{orth}} = \sum_{c} \|(\mathbf{U}_E^{(c)})^\top \mathbf{p}_{\text{mech}}^{(c)}\|_2^2
\label{eq:orth}
\end{equation}
which keeps the environment basis orthogonal to the mechanism prototype $\mathbf{p}_{\text{mech}}^{(c)}$ (the running mean of $\mathbf{z}_{\text{mech}}$ over all seen samples of event $c$, cf.\ Eq.~\ref{eq:zmech}). We define $\mathcal{L}_{\text{CIR}} = \mathcal{L}_{\text{inv}} + \mathcal{L}_{\text{equiv}}$, while $\mathcal{L}_{\text{orth}}$ is weighted separately (Section~\ref{sec:training}). CIR and $\mathcal{L}_{\text{orth}}$ regularize training; the projection layer deconfounds at inference.

\begin{figure*}[!t]
  \centering
  \includegraphics[width=\textwidth]{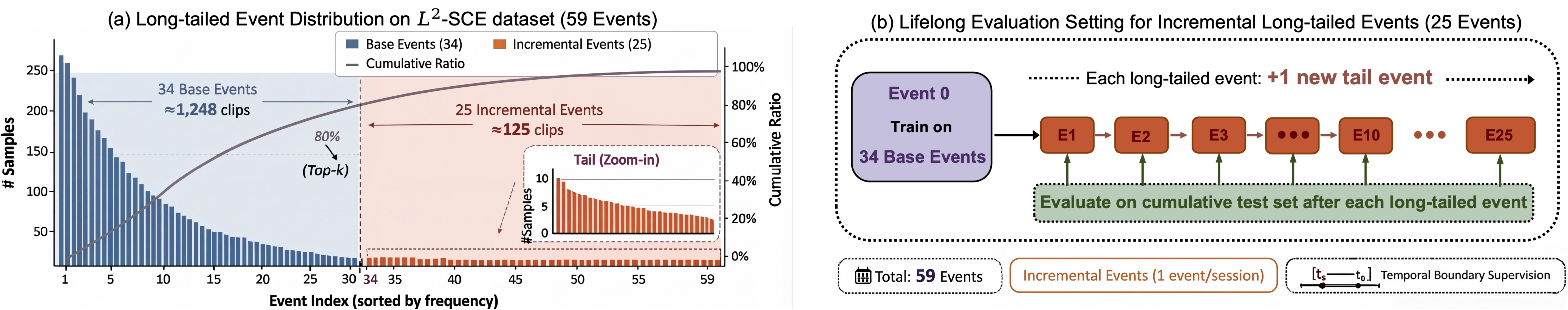}
  \caption{(a)~Category frequency distribution of \task dataset: 34 base events (blue, 1{,}248 samples) and 25 tail events (orange, 125 samples, rarest 1--2). (b)~Event-incremental evaluation setting with sequential tail-event arrival and cumulative evaluation.}
  \Description{(a) Bar chart of 59 security-critical event types sorted by frequency, showing a heavily skewed long-tailed distribution with base events in blue and tail events in orange. (b) Flowchart of the event-incremental evaluation setting.}
  \label{fig:dataset}
\end{figure*}

\subsection{Agentic Controller}
\label{sec:controller}

Since errors in generation or adaptation at one event can cascade into subsequent ones, the Agentic Controller validates each update through a closed-loop pipeline with self-reflective gating (Algorithm~\ref{alg:pipeline}).

\begin{algorithm}[t]
\caption{Agentic Controller Pipeline}
\label{alg:pipeline}
\begin{algorithmic}[1]
\REQUIRE Scarce samples $D_{\text{new}}^c$, model $f_\theta$, memory $\mathcal{B}$
\ENSURE Updated model $f_{\theta'}$, memory $\mathcal{B}'$
\FOR{round $r = 1$ \TO $R_{\max}$}
  \STATE CSG $\leftarrow$ \texttt{ExtractCSG}($D_{\text{new}}^c$, VLM) \COMMENT{Causal script}
  \STATE $\mathcal{C}_{\text{conf}} \leftarrow$ \texttt{Retrieve}(CSG, PrototypeMem)
  \STATE config $\leftarrow$ \texttt{Plan}(CSG, $\mathcal{C}_{\text{conf}}$, round=$r$) \COMMENT{Rule-based policy}
  \STATE $G_{\text{raw}} \leftarrow$ \texttt{Generate}(CSG, config, Director)
  \STATE $G^+, G^- \leftarrow$ \texttt{GCCG}($G_{\text{raw}}$, $f_\theta$) \COMMENT{Quality gate}
  \STATE $G^+_{\text{labeled}} \leftarrow$ \texttt{PseudoLabel}($G^+$, VLM) \COMMENT{Same backbone VLM}
  \STATE $f_{\theta^*} \leftarrow$ \texttt{Train}($D_{\text{new}}^c$, $G^+_{\text{labeled}}$, $\mathcal{B}$, $\mathcal{L}$)
  \STATE \texttt{CompatCalib}(adapter, $\mathcal{B}$, $f_\theta$, $f_{\theta^*}$)
  \STATE metrics $\leftarrow$ \texttt{Evaluate}($f_{\theta^*}$, all-seen val)
  \IF{old\_drop $\leq \delta_{\text{old}}$ \AND new\_gain $\geq \delta_{\text{new}}$}
    \STATE \texttt{Commit}($f_{\theta^*}$); \texttt{UpdateMemory}($\mathcal{B}$, $G^+$)
    \RETURN $f_{\theta^*}$, $\mathcal{B}'$
  \ELSE
    \STATE \texttt{Rollback}($f_\theta$); \texttt{AdjustPolicy}(config, metrics)
  \ENDIF
\ENDFOR
\RETURN $f_\theta$, $\mathcal{B}$ \COMMENT{No improvement; keep previous}
\end{algorithmic}
\end{algorithm}

\textbf{Self-Reflective Gating.} A checkpoint is committed only when the Macro-F1 decrease on base events (old\_drop) satisfies old\_drop~$\leq \delta_{\text{old}}$ and the new-event Macro-F1 gain (new\_gain) satisfies new\_gain $\geq \delta_{\text{new}}$; otherwise the controller rolls back and retries with adjusted policies.

\textbf{Adaptive policies.} Five rule groups govern adjustments: (1)~Old-event protection increases replay ratio; (2)~Generation quality adjusts curriculum and semantic thresholds; (3)~New-event learning boosts confusion samples; (4)~Temporal stability increases \doT ratio; and (5)~Projection health adjusts basis rank and $\lambda_{\text{orth}}$.

\subsection{Optimization}
\label{sec:training}

The total training loss has three parts:
\begin{equation}
\mathcal{L} = \underbrace{\mathcal{L}_{\text{task}}}_{\text{adaptation}} + \underbrace{\mathcal{L}_{\text{stab}}}_{\text{stability}} + \underbrace{\mathcal{L}_{\text{causal}}}_{\text{decoupling}}
\label{eq:total_loss}
\end{equation}

\noindent\textbf{Task adaptation.}
$\mathcal{L}_{\text{task}} = \mathcal{L}_{\text{sup}} + \lambda_{\text{bd}}\mathcal{L}_{\text{bd}}$,
where $\mathcal{L}_{\text{sup}}$ is cross-entropy classification plus Smooth L1 temporal regression on new-event samples, and $\mathcal{L}_{\text{bd}}$ is a boundary-shaping triplet loss (anchor: original, positive: same-mechanism variant, negative: \doM near-miss).

\noindent\textbf{Memory stability.}
$\mathcal{L}_{\text{stab}} = \lambda_{\text{rep}}\mathcal{L}_{\text{rep}} + \lambda_{\text{dist}}\mathcal{L}_{\text{dist}}$,
where $\mathcal{L}_{\text{rep}}$ replays stored exemplars from $\mathcal{B}$ with the same supervised objective and $\mathcal{L}_{\text{dist}}$ applies KL-divergence distillation from the frozen previous checkpoint~\cite{hinton2015distilling}.

\noindent\textbf{Causal decoupling.}
$\mathcal{L}_{\text{causal}} = \lambda_{\text{CIR}}\mathcal{L}_{\text{CIR}} + \lambda_{\text{orth}}\mathcal{L}_{\text{orth}}$
combines CIR (Eq.~\ref{eq:inv},~\ref{eq:equiv}) with the orthogonal penalty (Eq.~\ref{eq:orth}).

\noindent For synthetic samples, temporal regression is gated by $\mathbb{1}(S_{\text{temp}}^{(n)} > \tau_{\text{temp}})\cdot S_{\text{temp}}^{(n)}$, where $(n)$ indexes individual samples and $\tau_{\text{temp}}$ is the GCCG temporal threshold; demoted samples thus contribute no temporal gradient.

\section{Experimental Settings}
\label{sec:settings}

\subsection{Dataset Construction and Lifelong Setting}
\label{sec:dataset}

We construct the \task dataset based on the publicly available ECVA~\cite{du2024ecva}, which provides rich event types with a naturally long-tailed distribution, making it well-suited for simulating the \task incremental scenario. We merge semantically overlapping subcategories and filter ambiguous ones, yielding 59 security-critical event types with 1,373 annotated video clips, each labeled with an event class and temporal boundary $[t_s, t_e]$. We divide each event type into training, validation, and test sets with a 7:1:2 ratio. The resulting distribution exhibits a heavy long tail (Figure~\ref{fig:dataset}(a)): 34 base events $\mathcal{C}_0$ account for 1,248 samples, whereas the remaining 25 tail events $\mathcal{C}_1, \dots, \mathcal{C}_{25}$ share only 125 samples in total. We adopt a lifelong setting (Figure~\ref{fig:dataset}(b)): the base model is trained on $\mathcal{C}_0$ with full supervision, and the 25 tail events arrive sequentially, each with its naturally scarce samples. Evaluation is performed on the cumulative test set after each event. This distribution directly instantiates both challenges: extreme tail-event scarcity creates Intra-Event Scarcity, while semantically related incremental pairs share visual contexts that compound Inter-Event Interference.

\subsection{Baselines}
\label{sec:baselines}

We compare \method against 13 baselines across five categories. All methods share the same Qwen3-VL-8B-Instruct~\cite{bai2025qwen3vl} backbone with identical LoRA~\cite{hu2022lora} configuration. To isolate algorithmic differences, all LL baselines are further augmented with the same Wan2.2-generated synthetic data as \method (denoted w/ GenData in Table~\ref{tab:main}).

\noindent\textbf{Non-LL.} \textbf{Seq.~FT} performs naive sequential fine-tuning without forgetting mitigation. \textbf{Holmes-VAU}~\cite{zhang2025holmesvau} is a state-of-the-art SVU method, adapted with its security-critical event-focused temporal attention to our backbone.

\noindent\textbf{LL.} We evaluate regularization-based (\textbf{EWC}~\cite{kirkpatrick2017overcoming}, \textbf{LwF}~\cite{li2018learning}), replay-based (\textbf{Exp.~Replay}~\cite{rebuffi2017icarl}, \textbf{iCaRL}~\cite{rebuffi2017icarl}, \textbf{Replay+Distill}), and advanced hybrid methods (\textbf{DER++}~\cite{buzzega2020dark}, \textbf{BiC}~\cite{wu2019large}).

\noindent\textbf{Few-Shot LL.} \textbf{FACT}~\cite{zhou2022forward} and \textbf{TEEN}~\cite{wang2023teen} are few-shot event-incremental baselines; both freeze the backbone and use nearest-mean classification.

\noindent\textbf{Generative.} \textbf{GenReplay}~\cite{shin2017continual} uses the same Wan2.2 generator~\cite{wan2025wan} without GCCG or causal script guidance.

\noindent\textbf{Video LL.} \textbf{CCGS}~\cite{ccgs2024} is a state-of-the-art incremental few-shot video action recognition method, adapted with the same temporal localization head for direct comparison.

\begin{table*}[t]
\caption{Results on \task across 25 incremental events. Final metrics for DER++, BiC, CCGS, and \method are mean$\pm$std over five shuffled orders; others use the primary order. All methods use the same base model, and rows marked w/ GenData share the same synthetic data. Green/purple denote the full/ablated \method. MRE: Eq.~\ref{eq:mre}; ``--'': no memory. Best/second-best results are \textbf{bold}/\underline{underlined}.}
\label{tab:main}
\label{tab:ablation}
\resizebox{0.97\textwidth}{!}{%
\renewcommand{\arraystretch}{0.88}%
\setlength{\tabcolsep}{4pt}%
\begin{tabular}{l l cc cc cc cc ccc ccc}
\toprule
 & & \multicolumn{2}{c}{\textit{E5}} & \multicolumn{2}{c}{\textit{E10}} & \multicolumn{2}{c}{\textit{E15}} & \multicolumn{2}{c}{\textit{E20}} & \multicolumn{3}{c}{\textit{Final (E25)}} & \multicolumn{3}{c}{\textit{Stability}} \\
\cmidrule(lr){3-4} \cmidrule(lr){5-6} \cmidrule(lr){7-8} \cmidrule(lr){9-10} \cmidrule(lr){11-13} \cmidrule(lr){14-16}
Category & Method & F1$\uparrow$ & Acc$\uparrow$ & F1$\uparrow$ & Acc$\uparrow$ & F1$\uparrow$ & Acc$\uparrow$ & F1$\uparrow$ & Acc$\uparrow$ & F1$\uparrow$ & Acc$\uparrow$ & tIoU$\uparrow$ & Fgt.$\downarrow$ & BWT$\uparrow$ & MRE$\uparrow$ \\
\midrule
Non-LL & Seq.\ FT (Qwen3-VL-8B-Instruct) & 58.0 & 65.3 & 52.9 & 59.5 & 49.9 & 54.7 & 44.5 & 50.1 & 38.3 & 42.4 & 31.9 & 21.9 & $-$17.6 & -- \\
 & Holmes-VAU~\cite{zhang2025holmesvau} & 56.4 & 63.6 & 52.0 & 59.8 & 45.8 & 52.8 & 42.0 & 50.6 & 36.2 & 45.1 & 32.4 & 14.5 & $-$10.7 & -- \\
\specialrule{0.4pt}{1pt}{1pt}
\multirow{7}{*}{Lifelong Learning (LL)} & EWC~\cite{kirkpatrick2017overcoming} w/ GenData & 57.4 & 64.3 & 52.4 & 59.5 & 48.3 & 54.7 & 42.6 & 50.1 & 38.6 & 43.9 & 31.7 & 23.7 & $-$17.4 & -- \\
 & LwF~\cite{li2018learning} w/ GenData & 54.6 & 62.3 & 52.4 & 59.8 & 46.6 & 54.1 & 40.7 & 49.4 & 37.7 & 45.3 & 30.0 & 14.6 & $-$11.1 & -- \\
 & Exp.\ Replay~\cite{rebuffi2017icarl} w/ GenData & 61.5 & 67.9 & 55.8 & 60.1 & 52.2 & 53.9 & 44.1 & 42.3 & 41.9 & 38.1 & 26.1 & 21.2 & $-$2.1 & 7.6 \\
 & iCaRL~\cite{rebuffi2017icarl} w/ GenData & 51.0 & 54.9 & 51.3 & 54.2 & 45.9 & 48.0 & 43.7 & 48.1 & 36.7 & 36.2 & 28.1 & 20.3 & $-$7.1 & 6.9 \\
 & Replay+Distill w/ GenData & 59.2 & 67.2 & 56.4 & 63.8 & 48.2 & 56.3 & 42.0 & 52.1 & 38.1 & 48.4 & 27.3 & 16.5 & $-$16.5 & 10.5 \\
 & DER++~\cite{buzzega2020dark} w/ GenData & 61.2 & \underline{69.2} & \underline{58.2} & \underline{65.0} & 54.5 & \underline{61.3} & \underline{52.0} & \underline{57.9} & $\underline{48.6{\pm}0.6}$ & $\underline{55.9{\pm}0.9}$ & $\underline{32.1{\pm}1.2}$ & $17.8{\pm}0.8$ & $-13.8{\pm}2.2$ & \underline{11.8} \\
 & BiC~\cite{wu2019large} w/ GenData & 60.1 & 67.2 & 57.6 & 63.3 & \underline{55.8} & 58.4 & 51.7 & 51.4 & $45.8{\pm}1.1$ & $43.6{\pm}1.5$ & $27.9{\pm}2.3$ & $\underline{9.8{\pm}1.8}$ & $\underline{8.2{\pm}2.0}$ & 10.8 \\
\specialrule{0.4pt}{1pt}{1pt}
\multirow{2}{*}{Few-Shot LL} & FACT~\cite{zhou2022forward} & 51.0 & 55.2 & 50.4 & 53.4 & 46.4 & 48.8 & 43.1 & 46.9 & 37.5 & 37.4 & 24.7 & 21.2 & $-$7.1 & 7.8 \\
 & TEEN~\cite{wang2023teen} & 50.5 & 54.5 & 51.8 & 52.8 & 45.0 & 47.2 & 42.3 & 44.1 & 38.8 & 38.8 & 24.5 & 29.3 & $-$27.0 & 11.7 \\
\specialrule{0.4pt}{1pt}{1pt}
Generative & GenReplay~\cite{shin2017continual} & 61.5 & 67.5 & 53.4 & 53.9 & 49.0 & 46.4 & 43.4 & 37.5 & 39.2 & 34.5 & 24.0 & 28.7 & $-$27.9 & 5.9 \\
Video LL & CCGS~\cite{ccgs2024} & \underline{63.3} & 68.8 & 55.9 & 60.1 & 53.4 & 54.1 & 46.2 & 42.8 & $37.9{\pm}1.2$ & $33.5{\pm}0.9$ & $29.2{\pm}2.0$ & $25.9{\pm}2.8$ & $-17.5{\pm}1.1$ & 6.1 \\
\midrule
\rowcolor{bestbg}
\textbf{Ours} & \textbf{\method (Full)} & \textbf{64.0} & \textbf{71.4} & \textbf{60.3} & \textbf{67.9} & \textbf{56.1} & \textbf{61.6} & \textbf{54.5} & \textbf{59.9} & $\mathbf{51.5{\pm}0.9}$ & $\mathbf{56.7{\pm}0.7}$ & $\mathbf{34.4{\pm}1.1}$ & $\mathbf{4.0{\pm}0.6}$ & $\mathbf{23.2{\pm}1.6}$ & \textbf{13.3} \\
\rowcolor{ablationbg}
 & \quad w/o CVG & 59.9 & 67.2 & 57.0 & 64.1 & 52.6 & 59.2 & 51.9 & 58.9 & 47.2 & 54.4 & 33.3 & 5.2 & $+$18.4 & 12.6 \\
\rowcolor{ablationbg}
 & \quad w/o GCCG (ungated) & 62.6 & 70.1 & 59.4 & 67.3 & 57.2 & 61.3 & 53.9 & 57.4 & 46.9 & 53.7 & 29.7 & 14.6 & $-$7.3 & 10.8 \\
\rowcolor{ablationbg}
 & \quad w/o CIR ($\mathcal{L}_{\text{inv}}{+}\mathcal{L}_{\text{equiv}}$) & 57.4 & 67.5 & 54.7 & 63.0 & 52.1 & 61.1 & 49.1 & 58.4 & 46.1 & 57.1 & 28.5 & 6.6 & $-$2.9 & 13.0 \\
\rowcolor{ablationbg}
 & \quad w/o Causal Proj. & 59.3 & 67.5 & 56.6 & 64.7 & 51.9 & 60.5 & 51.1 & 58.7 & 48.8 & 56.1 & 25.9 & 3.6 & $+$13.8 & 13.3 \\
\rowcolor{ablationbg}
 & \quad w/o Compat.\ Adapter & 60.8 & 68.5 & 58.9 & 65.6 & 53.1 & 58.7 & 52.8 & 58.2 & 46.8 & 54.0 & 24.0 & 10.6 & $+$3.2 & 11.6 \\
\rowcolor{ablationbg}
 & \quad w/o Agentic Controller & 61.9 & 69.2 & 60.2 & 67.1 & 55.3 & 60.5 & 53.1 & 58.9 & 50.6 & 56.8 & 26.9 & 6.4 & $+$7.9 & 13.0 \\
\bottomrule
\end{tabular}%
}
\end{table*}

\begin{figure*}[!t]
  \centering
  \includegraphics[width=\textwidth]{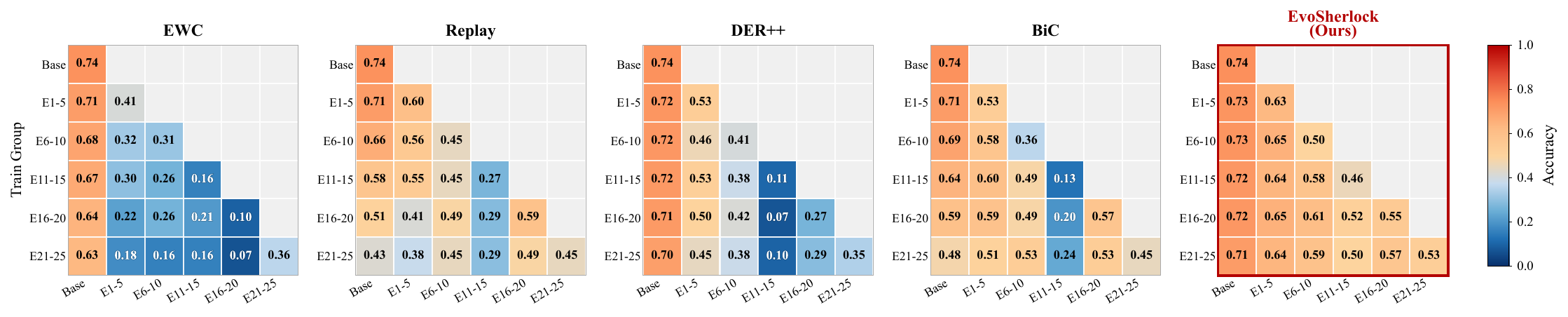}
  \caption{Cross-event forgetting heatmaps. Baselines show pronounced off-diagonal degradation, whereas \method maintains consistently higher accuracy.}
  \Description{Five heatmaps: baselines show increasing off-diagonal degradation while EvoSherlock maintains consistently high accuracy across all event combinations.}
  \label{fig:forgetting_heatmap}
\end{figure*}

\begin{figure*}[!t]
  \centering
  \includegraphics[width=\textwidth]{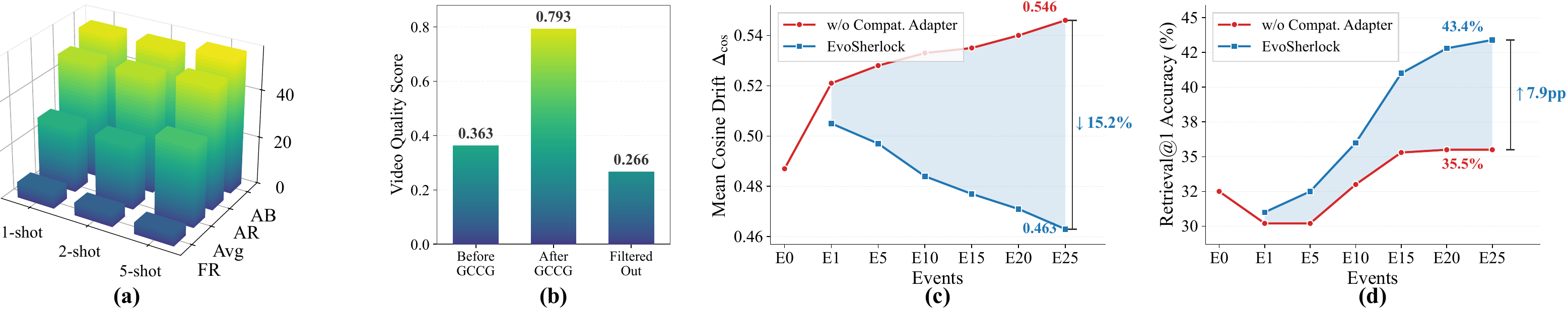}
  \caption{Data scarcity, generation quality, and prototype drift. (a)~1/2/5-shot performance. (b)~GCCG quality scores. (c)~Prototype drift and (d)~old-event Retrieval@1 with and without the Compatibility Adapter.}
  \Description{Four-panel figure: (a) 3D bar chart of 1/2/5-samples accuracy with Forgetting Rate; (b) quality score distributions under GCCG gating; (c) line chart of prototype cosine drift; (d) line chart of old-class retrieval accuracy.}
  \label{fig:shot_ablation}
  \label{fig:prototype_drift}
\end{figure*}

\begin{figure*}[!t]
  \centering
  \includegraphics[width=0.98\textwidth]{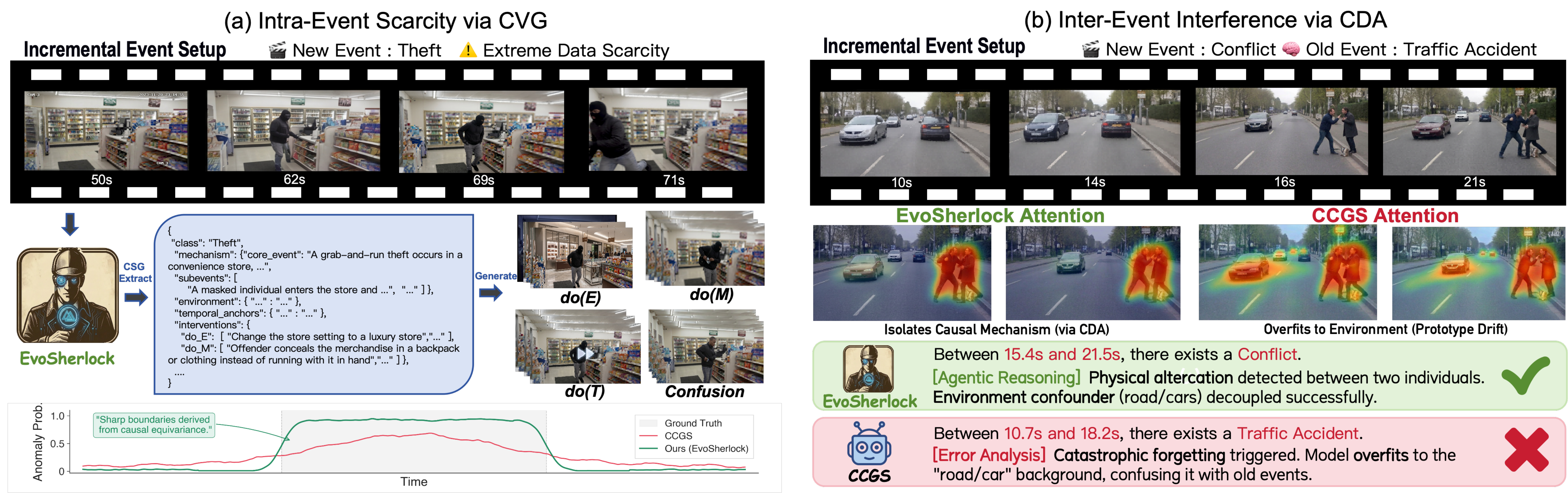}
  \caption{Qualitative examples. (a)~CVG generates causally grounded counterfactuals and sharper temporal boundaries. (b)~CDA attends to event mechanisms, whereas CCGS overfits to shared context.}
  \Description{Two-part case study: (a) CVG pipeline showing CSG extraction, counterfactual generation, and temporal localization comparison; (b) CDA attention heatmaps contrasting mechanism-focused vs.\ environment-focused predictions.}
  \label{fig:case_study}
\end{figure*}

\subsection{Evaluation Metrics}
\label{sec:metrics}

For the \task setting, we adopt several evaluation metrics. For event classification, we report Macro-F1 and Top-1 Accuracy, where Macro-F1 is the primary metric due to event imbalance across events. For temporal localization, we use tIoU mAP at thresholds of 0.3 and 0.5. For lifelong learning, we report Forgetting and Backward Transfer (BWT), where lower Forgetting indicates better retention and positive BWT indicates beneficial backward transfer.

For memory retention efficiency, we define a new metric, MRE, to jointly assess retention quality and memory cost. Let $\bar{A}_T$ denote the final-event accuracy, $A_0$ the base-event accuracy, $B_{\text{mem}}$ the number of stored memory items (\eg exemplars, prototypes, and causal scripts), and $|\mathcal{C}_{\leq T}|$ the number of observed events up to event $T$:
\begin{equation}
\text{MRE} =
\frac{\bar{A}_T \cdot \left(1 - \tfrac{\text{Fgt}}{A_0}\right)}
{1 + \log_2\!\left(1 + \tfrac{B_{\text{mem}}}{|\mathcal{C}_{\leq T}|}\right)}
\label{eq:mre}
\end{equation}
where the numerator captures retained performance after penalizing forgetting, and the denominator penalizes memory usage per event. Higher MRE indicates better efficiency.

\subsection{Implementation Details}
\label{sec:impl}

Since the above baselines employ different experimental settings, we reimplement these approaches using their publicly released code on our \task dataset, ensuring a fair and comprehensive comparison. We retain their published settings and tune the remaining hyperparameters on the validation set. Specifically, \method uses AdamW~\cite{loshchilov2019decoupled} with cosine annealing, a learning rate of $3\times10^{-6}$, batch size 2, and 3 epochs per event. LoRA targets \texttt{q/k/v/o\_proj} with rank 16, $\alpha\!=\!32$, and dropout 0.25. The causal projection uses rank $k\!=\!8$ and a learning rate of $10^{-4}$. We set $(\lambda_{\text{dist}},\lambda_{\text{bd}},\lambda_{\text{CIR}},\lambda_{\text{orth}})=(4.0,1.5,0.5,0.5)$ and $T\!=\!2.0$. GCCG uses $(w_1,w_2,w_3,w_4)=(0.3,0.2,0.2,0.3)$, $\tau_{\text{sem}}\!=\!\tau_{\text{temp}}\!=\!0.5$, $\delta_{\text{old}}\!=\!0.05$, and $\delta_{\text{new}}\!=\!0.1$. Experiments use bf16 on two NVIDIA RTX 4090 GPUs; the 25-event pipeline takes approximately 30 hours, including 62\% for video generation. All methods use the primary event order for trajectory comparisons; DER++, BiC, CCGS, and \method are additionally evaluated over five independently shuffled orders, with final metrics reported as mean and standard deviation in Table~\ref{tab:main}.

\section{Results and Discussion}
\label{sec:results}

\subsection{Experimental Results}
\label{sec:main_results}

\textbf{Existing paradigms remain insufficient for {\boldmath$L^2$}-SCE.} Holmes-VAU~\cite{zhang2025holmesvau} yields competitive tIoU but the lowest final F1; CCGS~\cite{ccgs2024} incurs 21.9 more Forgetting than \method; GenReplay (the same Wan2.2 generator, without GCCG) has the worst BWT; and FSCIL methods (FACT, TEEN) perform poorly. These results confirm that \task requires causal grounding and backbone adaptation (Figure~\ref{fig:forgetting_heatmap}).

\textbf{\method consistently outperforms all baselines in classification, temporal localization, and lifelong stability.} In Table~\ref{tab:main}, \method surpasses DER++ by 2.9 final Macro-F1 and 0.8 Acc. For stability, \method reduces Forgetting by 5.8 over BiC and improves BWT by 15.0. MRE (Eq.~\ref{eq:mre}) further confirms superior memory efficiency, exceeding DER++ by over 12\%. These gains validate CVG's and CDA's challenge-driven designs.

\textbf{\method achieves the highest temporal localization through causally grounded synthesis.} Across five event orders, \method outperforms DER++ by 2.3 in mean tIoU@0.5; under the primary event order, it outperforms Holmes-VAU by 1.4. These gains arise from GCCG filtering temporally inconsistent synthetic samples and the \doT equivariance loss (Eq.~\ref{eq:equiv}) providing direct boundary training signals. This demonstrates that causal generation not only improves classification but also preserves precise temporal boundaries under the event-incremental setting.

\subsection{Effectiveness of Video Generation}
\label{sec:ablation_cvg}

\noindent\textbf{CVG contributes primarily to new-event adaptation.} Disabling CVG degrades F1 by 4.1 with minimal impact on Forgetting and BWT, confirming its targeted role in augmenting scarce tail-event training data through causally grounded counterfactual synthesis.

\noindent\textbf{Removing GCCG is worse than removing generation entirely.} Ungated generation drops F1 below the no-CVG variant and raises Forgetting by 10.4 with a BWT collapse of over 30 points, confirming GCCG is essential to prevent noisy labels from harming stability. As shown in Figure~\ref{fig:shot_ablation}(b), After GCCG samples achieve higher video quality scores than Before GCCG and Filtered Out ones, validating that GCCG separates causally faithful videos from corrupted ones.

\subsection{Effectiveness of Decoupling and Alignment}
\label{sec:ablation_cda}

\noindent\textbf{CIR yields the largest single-component impact.} Removing CIR produces the largest F1 drop and a BWT swing of $-$25.7 into negative territory, confirming that CIR is essential for preventing cross-event feature entanglement.

\noindent\textbf{CDA sub-components are complementary.} Removing the Causal Projection layer degrades F1 and tIoU while Forgetting remains comparable, indicating that deconfounding primarily boosts discriminability. Removing the Compatibility Adapter increases Forgetting by 6.4 and reduces BWT by 19.6, as prototype drift erodes retrieval consistency. Together, the two sub-components jointly address Inter-Event Interference from complementary perspectives.

\subsection{Effectiveness of Agentic Controller}
\label{sec:ablation_ctrl}

\noindent\textbf{Agentic Controller prevents temporal localization collapse.} Removing the Agentic Controller reduces tIoU by 6.9 and BWT by 14.9 with minimal F1 impact ($-$0.7), indicating the controller's primary role is rejecting low-quality updates that corrupt temporal boundaries. This confirms that self-reflection is critical for maintaining temporal precision under the event-incremental setting.

\subsection{Qualitative Analysis}
\label{sec:qualitative}

\textbf{Robustness to data scarcity.}
Figure~\ref{fig:shot_ablation}(a) shows results with 1, 2, and 5 samples per event. \method remains competitive across all settings, and Forgetting decreases as samples increase, confirming CVG's robustness to scarcity. Even with only 1 sample, \method outperforms baselines trained with full data, highlighting the effectiveness of causal counterfactual augmentation.

\textbf{Prototype drift quantification.}
To verify that the Compatibility Adapter mitigates the prototype drift of Inter-Event Interference, Figure~\ref{fig:shot_ablation}(c) shows monotonic drift without it and 15.2\% reduction at E25 with it. Figure~\ref{fig:shot_ablation}(d) shows 7.9\,pp higher retrieval accuracy at E25, linking prototype stability to the BWT advantage.

\textbf{CVG generation and temporal localization.}
As shown in Figure~\ref{fig:case_study}(a), \method extracts a CSG from a ``Theft'' sample and synthesizes counterfactual videos along four axes. The localization curve shows sharper boundaries than CCGS and closer ground-truth alignment.

\textbf{Causal projection effect.}
Figure~\ref{fig:case_study}(b) shows CDA isolating the causal mechanism from road/car confounders in ``Conflict'' \vs ``Traffic Accident''; CCGS misclassifies via background overfitting, confirming causal decoupling prevents inter-event interference.

\section{Conclusion}
\label{sec:conclusion}

In this paper, we formalize a new \task task and propose \method, a causal-enhanced approach consisting of the CVG and CDA modules and an Agentic Controller, to effectively address the Intra-Event Scarcity and Inter-Event Interference challenges, respectively. Extensive experiments on our constructed \task dataset demonstrate that \method substantially outperforms several advanced baselines in classifying and temporally localizing security-critical events, achieving positive backward transfer while maintaining strong new-event adaptation. The constructed \task benchmark further provides a standardized testbed for advancing lifelong security-oriented video understanding.

In future work, we will integrate agentic skill-learning frameworks such as Memento-Skills~\cite{zhou2026memento} and OpenClaw-RL~\cite{wang2026openclaw}, enabling the controller to evolve task-specific skills with less manual prompt engineering. We will also extend \method to additional security-critical scenarios, including drone surveillance and autonomous driving, and evaluate cross-dataset generalization with larger event taxonomies.

\begin{acks}
This work was supported by three NSFC grants, i.e., No. 62576234, No.62376178, No.62376182 and sponsored by CIPS-LMG Huawei Innovation Fund. This work was also supported by Collaborative Innovation Center of Novel Software Technology and Industrialization, Jiangsu Province Talent Programme Qinglan Project and a Project Funded by the Priority Academic Program Development of Jiangsu Higher Education Institutions (PAPD).
\end{acks}

\bibliographystyle{ACM-Reference-Format}
\balance
\bibliography{references}

\end{document}